\documentclass{article}

 \usepackage[preprint]{neurips_2026}

\usepackage[utf8]{inputenc} 
\usepackage[T1]{fontenc}    
\usepackage{hyperref}       
\usepackage{url}            
\usepackage{booktabs}       
\usepackage{amsfonts}       
\usepackage{nicefrac}       
\usepackage{microtype}      
\usepackage{xcolor}         
\usepackage{amsmath,amssymb}
\usepackage{algorithm}
\usepackage{algorithmic}
\usepackage{comment}
\usepackage{graphicx}
\usepackage{color}
\usepackage{subcaption}
\usepackage{amsmath,amsfonts,bm}

\def\eqref#1{equation~\ref{#1}}

\def\1{\bm{1}}

\DeclareMathAlphabet{\mathsfit}{\encodingdefault}{\sfdefault}{m}{sl}
\SetMathAlphabet{\mathsfit}{bold}{\encodingdefault}{\sfdefault}{bx}{n}

\title{Dynamic Representational Synchrony through Collective Predictive Coding: A Computational Model of Parent--Infant Homeostatic Co-Regulation}

\author{%
  Yushi Tsubamoto\\
  Graduate School of Medicine\\
  The University of Osaka\\
  565-0871, Yamadaoka, Suita, Osaka\\
  \texttt{u935741h@alumni.osaka-u.ac.jp} \\
  \And
  Takato Horii \\
  Graduate School of Engineering Science\\
  The University of Osaka\\
  560-0043, Machikaneyama, Toyonaka Osaka\\
  \texttt{takato@sys.es.osaka-u.ac.jp}
}

\begin{document}
\maketitle

\begin{abstract}
Inter-brain synchrony (IBS) observed in real-time dyadic interactions, including parent--infant exchanges, suggests that two agents can align their internal representations through interaction.
Yet computational accounts of how such alignment can arise between agents that have only local sensory access and asymmetric internal knowledge remain underdeveloped.
We propose a constructive model of parent--infant homeostatic co-regulation that integrates a POMDP formulation of active interoceptive inference with the Metropolis--Hastings Naming Game (MHNG) derived from the Collective Predictive Coding (CPC) hypothesis.
In our model, the parent and infant agents agree on homeostatic regulatory actions for the infant's visceral state through a shared communicative variable generated by a locally computable Metropolis--Hastings probability.
The parent observes the infant through body-generated exteroceptive cues, whereas the infant directly senses its own visceral state through interoception.
This difference in access modality is implemented as asymmetric generative-model knowledge: the parent knows how actions transform visceral states but must learn what the infant's bodily cues indicate, whereas the infant perceives its visceral state directly but must learn how actions affect it.
We quantify the degree of representational alignment using the Jensen--Shannon divergence between the two agents' latent representations. 
In a $6 \times 6$ visceral-state grid world, MHNG-mediated interaction regulated the infant's visceral state more adaptively than one-sided control conditions. Moreover, the agents’ latent representations not only became rapidly aligned but also remained dynamically coupled across successive state transitions, demonstrating dynamic latent-representation synchrony.
Notably, this synchrony emerged far earlier than the generative-model convergence and was maintained despite heterogeneous generative-model knowledge, indicating that it does not require fully shared world models.
These findings support CPC as a candidate computational framework for explaining how dynamic 
representational synchrony relevant to IBS can emerge through local interactions.
\end{abstract}

\section{Introduction}

Social interaction is accompanied by coordinated neural, behavioral and physiological dynamics across individuals\cite{schilbach2024synchrony,feldman2012parent}.
Hyperscanning studies have repeatedly reported inter-brain synchrony (IBS) during real-time social interaction, including parent–child interaction\cite{schilbach2024synchrony,hamilton2021hyperscanning}. Despite accumulating empirical evidence, however, theoretical computational accounts of how such interpersonal alignment emerges from dyadic interaction remain underdeveloped.
A key challenge is to explain how two agents, each having only local sensory access and private internal models, can nevertheless align their internal representations during interaction. 
In this paper, we treat the rapid alignment and sustained dynamic coupling of agents’ latent representations as a computational analogue of interpersonal synchrony reflected in IBS.

Parent–infant interaction provides a suitable domain for testing this idea because it involves not only information exchange but also regulation of bodily states and the learning of social capacities that support later socio-emotional development\cite{feldman2012parent}. 
The theory of constructed emotion emphasizes that interoception and socially learned categories play vital roles in emotion development\cite{hoemann2019emotion}.
Because infants cannot survive alone and depend on parents to regulate their ongoing physiology, early social learning may be grounded in the need to manage bodily states through interaction with others\cite{atzil2018growing}.
Notably, it is reported that greater mother–infant IBS is associated with enhanced emotion regulation abilities in children\cite{reindl2018brain}.
Here, we focus on a more minimal precursor: dyadic regulation of bodily states, socially mediated signals and rapid alignment of latent representations through interaction. 
Such regulation is inherently asymmetric in that parents may have better knowledge about regulatory actions, whereas infants have more direct access to their own bodily states. 
Therefore, a computational model of parent–infant homeostatic co-regulation should explain how agents with asymmetric knowledge and only locally available sensory access can coordinate to regulate bodily states while aligning their latent representations.

The free-energy principle and active inference provide a computational account of how biological agents maintain adaptive states by inferring the hidden causes of their sensory signals and selecting actions expected to realize preferred outcomes\cite{friston2010free}. 
In the context of interoception, active inference treats bodily regulation as a process in which an agent infers its bodily state and acts to reduce expected deviations from preferred physiological states\cite{seth2016active}.
Active-inference approaches to second-person neuroscience extend this idea to social interaction by modeling interpersonal coordination as reciprocal prediction and prediction-error minimization \cite{mayo2024dynamic, lehmann2024active}. 
Yet, a remaining challenge is to formulate social interaction in a way that does not require agents to directly access each other’s observations or to maintain deeply recursive models of the other, while still allowing their latent representations to become mutually aligned.
Thus, a computational account of interpersonal alignment requires a locally computable mechanism through which the beliefs of two agents can become mutually constrained.

The Collective Predictive Coding (CPC) hypothesis provides such a mechanism by treating communication as a process in which a shared variable constrains the posterior beliefs of multiple agents\cite{taniguchi2024collective}. 
CPC extends individual-level predictive coding to a multi-agent setting at the social level by formulating social interaction as decentralized Bayesian inference.
In this view, symbols should be understood broadly as a shared communicative variable including not only  a fully developed linguistic symbol but a cue, a label or an interactional signal.
The Metropolis--Hastings Naming Game (MHNG) offers an algorithmic implementation of this hypothesis\cite{taniguchi2023emergent, ebara2023multi}. 
In MHNG, one agent, the speaker, proposes a communicative sign, and the other agent, the listener, accepts or rejects it according to a Metropolis--Hastings acceptance probability that reflects how well the sign fits the listener's own generative model.
Because this accept--reject decision is computed from the listener's own model, the interaction can constrain both agents' beliefs without requiring either agent to observe the other's internal states directly. 
Empirical studies suggest that human communicative acceptance behavior can be approximated as MHNG, supporting the view that human communication can be interpreted within the CPC framework \cite{okumura2023metropolis, okumura2025co}.

We develop a constructive model of parent--infant homeostatic co-regulation by integrating a POMDP model of individual-level active interoceptive inference\cite{Heins2022,da2020active} with MHNG based on the CPC hypothesis\cite{taniguchi2023emergent, ebara2023multi}. 
In the proposed model, a shared communicative sign mediates agreement over regulatory actions for the infant's bodily state, without requiring either agent to directly access the other's private internal states. 
We ask whether local MHNG-mediated interaction can induce dynamic latent-representation synchrony, that is, whether two agents can become aligned and remain dynamically coupled despite asymmetric knowledge.
The main contributions are as follows:
\begin{enumerate}
    \item We formalize parent--infant homeostatic co-regulation as MHNG interaction between two active-interoceptive-inference agents with asymmetric sensory access and asymmetric generative-model knowledge.

     \item We empirically demonstrate that MHNG-mediated interaction regulates the infant's bodily state more adaptively than one-sided control conditions.

    \item We show that the two agents’ latent representations become aligned rapidly and remain  dynamically coupled across subsequent action-induced state transitions, even before their learned generative models converge. This provides a minimal constructive account of dynamic latent-representation synchrony under asymmetric sensory access and generative-model knowledge.
\end{enumerate}

\begin{figure}[t]
    \centering
    \includegraphics[width=0.5\linewidth]{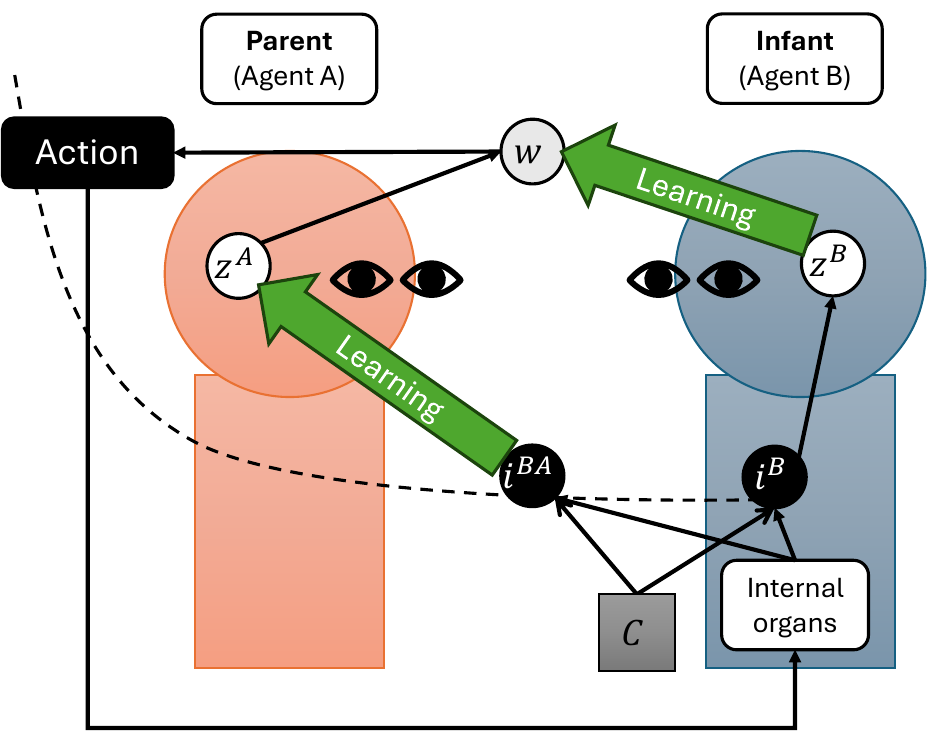}
    \caption{Graphical model of the proposed method.}
    \label{fig:graphical_model}
\end{figure}
 
\begin{figure}[t]
    \centering
    \includegraphics[width=0.9\linewidth]{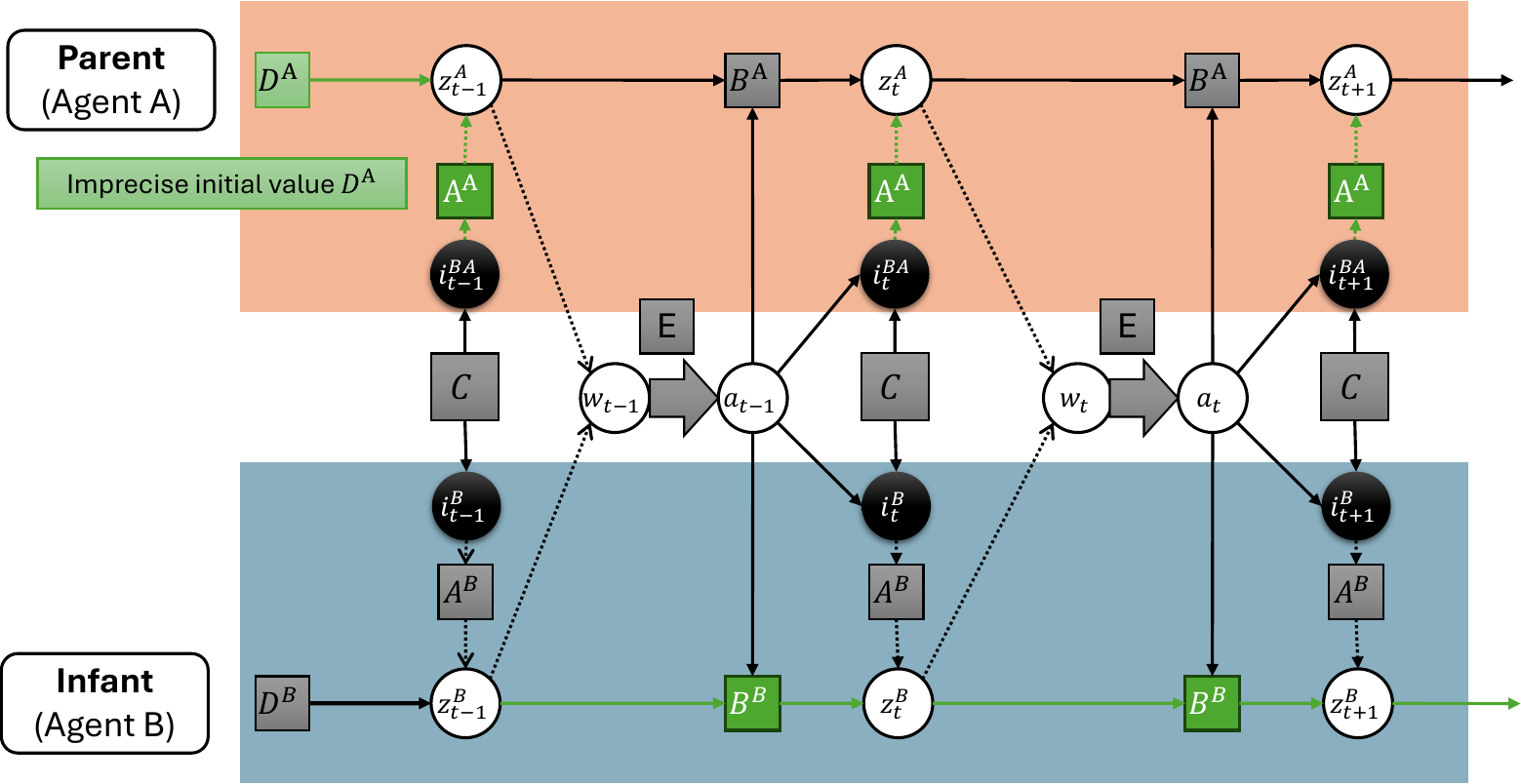}
    \caption{Graphical model expanded along the time axis.}
    \label{fig:time_series}
\end{figure}

\section{Proposed Model}
\subsection{Agent: POMDP Model}

Figures~\ref{fig:graphical_model} and~\ref{fig:time_series} show the graphical models of the proposed method. Imprecise parameters that need to be learned are highlighted in green. For agent $X \in \{A, B\}$, $z^{X}$ is a latent variable representing the agent's internal representation, and $w$ denotes the symbol generated between the agents. The proposed model assumes that two agents determine an action $a$ from the shared symbol $w$ and manage the visceral state through cooperative behavior.

The two agents are intentionally asymmetric in which components of the generative model they possess accurately and which components they must learn from interaction.
The parent agent has an accurate state-transition matrix $B^{A}$, encoding how regulatory actions transform visceral states, but must learn the sensory-generation matrix $A^{A}$ that links its inference of the infant's state to the observed exteroceptive cue $i^{BA}_{t}$. Conversely, the infant agent has an accurate sensory-generation matrix $A^{B}$ for direct interoceptive perception of its own visceral state but must learn the state-transition matrix $B^{B}$ that captures how actions affect its body.
This asymmetry, formalized in Section~\ref{subsec:online_learning}, reflects the everyday situation in which caregivers know how to regulate visceral states while infants more directly experience them.

\begin{align}
z^{X}_{t} &\sim P\!\left(z^{X}_{t} \mid z^{X}_{t-1}, a_{t-1}, B^{X}\right) \\
i^{X}_{t} &\sim P\!\left(i^{X}_{t} \mid z^{X}_{t}, A^{X}\right) \\
w_{t} &\sim P\!\left(w_{t} \mid z^{A}_{t}, z^{B}_{t}\right) \\
a_{t} &\sim P\!\left(a_{t} \mid w_{t}, E\right)
\end{align}

For the infant agent, $i^{X}_{t}$ corresponds to the interoceptive perception $i^{B}_{t}$, whereas for the parent agent it corresponds to the exteroceptive cue $i^{BA}_{t}$ generated by the infant agent's body.
For example, when the infant agent is in a low-energy state, it may cry, show rooting behavior, or make sucking-related movements. These bodily cues do not give the parent direct access to the infant's bodily state, but they serve as external evidence from which the parent agent can infer the infant's bodily state through vision, audition, touch, and other modalities.
For simplicity, the infant’s interoceptive observation $i^{B}$ and the parent’s exteroceptive cue $i^{BA}$ are encoded in the same one-hot state space. Thus, the asymmetry in the present model lies not in the amount of sensory information available to each agent, but in the access modality and in the agents’ generative-model knowledge: the infant has an accurate $A^B$, whereas the parent must learn $A^A$.
Each agent communicates so as to minimize expected free energy, and learns the matrices $A^{A}$ and $B^{B}$.
 
\subsection{Communication: MHNG}

We model symbol-mediated interaction using MHNG.
As an intuitive example, an infant acting as the speaker may propose a feeding-related sign such as ``num-num.''
If the parent, acting as the listener, infers that sleeping is currently more appropriate, the parent may reject this proposal according to the Metropolis--Hastings acceptance probability and instead share a sleep-related sign such as ``night-night.''
The shared sign is mapped to the cooperative regulatory action such as putting the infant to sleep.

The target distribution $P(w)$ is formulated using a Product-of-Experts (PoE) approximation as follows\cite{ebara2023multi}:
\begin{equation}
P(w) = P\!\left(w \mid z^{A}, z^{B}\right) \propto P\!\left(w \mid z^{A}\right) P\!\left(w \mid z^{B}\right).
\end{equation}
 
In this model, the posterior distribution of the symbol $w$ given the Agent X's latent representation $z^X$ is computed based on the expected free energy $F^X_{\text{exp}}$ through the following procedure.
 
\begin{enumerate}
\item Compute the expected free energy for each action $a$:
\begin{equation}
F^X_{\text{exp}}(a) =\mathbb{E}_{q^X}\!\left[H\!\left[p(i^X \mid z^X)\right]\right] + D_{\text{KL}}\!\left[q(i^X \mid a) \,\|\, p(i^X \mid C)\right].
\end{equation}

\item Convert this to the expected free energy for each symbol by taking the inner product with the interpretation matrix $E = P(a \mid w)$:
\begin{equation}
G^X(w) = \sum_{a} P(a \mid w)\, F^X_{\text{exp}}(a).
\end{equation}
 
\item Obtain the posterior distribution from the per-symbol expected free energy $G(w)$:
\begin{equation}
P(w \mid z^X) = \mathrm{softmax}\!\left(-G^X(w)\right) = \frac{\exp\!\left(-G^X(w)\right)}{\sum_{w'} \exp\!\left(-G^X(w')\right)}.
\end{equation}
\end{enumerate}
 
By the above, an agent stochastically selects a symbol based on its internal states. Here, the prior preference $C$ is a prior distribution that represents the desirability of an observation $i^{X}$. We assume that it is biologically shared, and both agents possess common parameters. The interpretation matrix $E$ is a probability distribution that represents the relationship between symbols $w$ and actions $a$. In this experiment, we assume a one-to-one correspondence between symbols and actions, and use the identity matrix.
 
Because the speaker proposes a symbol based on its own internal states, the proposal distribution is given by
\begin{equation}
Q(w' \mid w) = P\!\left(w' \mid z^{Sp}\right).
\end{equation}
 
The probability that the listener accepts the proposal $w'$ is, by the Metropolis--Hastings method,
\begin{equation}
r^{MH} = \min\!\left(1, \, \frac{P(w')\, Q(w \mid w')}{P(w)\, Q(w' \mid w)}\right) = \min\!\left(1, \, \frac{P(w' \mid z^{Li})}{P(w \mid z^{Li})}\right).
\end{equation}
 
The acceptance probability of $w'$ proposed by the speaker can be computed using only the listener's parameters, without knowing the partner's internal states.
Both agents alternately take the speaker and listener roles within one iteration.
 
\subsection{Online Learning of the Generative Model}\label{subsec:online_learning}
 
In this model, the parent and infant agents learn the generative model online following a Dirichlet distribution through their interaction\cite{da2020active}. The parent agent is assumed to possess an accurate state-transition matrix $B = P(z_{t+1} \mid z_{t}, a_{t})$, while the sensory generation matrix $A = P(i \mid z)$ needs to be learned. This models a situation in which the parent knows how to manage visceral states (e.g., that eating fills the stomach) but does not know the infant's current visceral state (e.g., whether the infant is hungry). Conversely, the infant agent is assumed to possess an accurate sensory generation matrix $A$, while the state-transition matrix $B$ needs to be learned. This models a situation in which the infant can perceive its current visceral state through interoceptive sensory signals, but does not know how to maintain a comfortable visceral state (e.g., whether eating or covering oneself with a blanket would be more comfortable). In this way, the present paper models the asymmetry of knowledge that exists in real parent--infant relationships.
 
\subsubsection{Parent Agent: Learning the $A$ Matrix}
 
The sensory generation matrix $A^{A}$ of the parent agent (Agent A) represents the correspondence between the latent representation $z^{A}_{t}$ of the other-model that predicts the infant's visceral state, and the exteroceptive sensory signal $i^{BA}_{t}$ generated by the infant's body. We model $A$ as being generated from a Dirichlet parameter $\alpha_{A}$, and learn it as follows.
 
\begin{enumerate}
\item \textbf{Setting the prior.} Assuming no prior knowledge about the relationship between $z^{A}$ and the resulting observation $i^{BA}$, we use a uniform Dirichlet distribution over observations as the prior.
 
\item \textbf{Online parameter learning.} At each step $t$, we observe the exteroceptive sensory signal $i^{BA}_{t}$ generated by the infant and infer the posterior $q(z^{A}_{t})$. The corresponding parameter $\alpha_{A}(\cdot, i^{BA}_{t})$ is updated as
\begin{equation}
\alpha_{A}^{\text{new}}(\cdot, i^{BA}_{t}) = \alpha_{A}^{\text{old}}(\cdot, i^{BA}_{t}) + q(z^{A}_{t}).
\end{equation}
 
\item \textbf{Computing the posterior.} Using the updated parameters $\alpha_{A}^{\text{new}}$, we compute the posterior of the $A$ matrix and apply it from the next step onward.
\end{enumerate}
 
\subsubsection{Infant Agent: Learning the $B$ Matrix}
 
The state-transition matrix $B^{B}$ of the infant agent (Agent B) represents the correspondence between the cooperative action $a_{t-1}$ of parent and infant, and the latent representation $z^{B}_{t}$ that predicts its own interoceptive sensory signals. We model $B$ as being generated from a Dirichlet parameter $\beta_{B}$, and learn it as follows.
 
\begin{enumerate}
\item \textbf{Setting the prior.} Assuming no prior knowledge about the transitional relationship between an action $a$ and the resulting observation $i^{B}$, we use a uniform Dirichlet distribution as the prior.
 
\item \textbf{Online parameter learning.} At each step, we infer the posteriors $q(z^{B}_{t-1})$ and $q(z^{B}_{t})$, and update the parameter $\beta_{B}(\cdot, \cdot, a_{t-1})$ corresponding to action $a_{t-1}$ by computing the expected value via the outer product:
\begin{equation}
\beta_{B}^{\text{new}}(\cdot, \cdot, a_{t-1}) = \beta_{B}^{\text{old}}(\cdot, \cdot, a_{t-1}) + q(z^{B}_{t}) \otimes q(z^{B}_{t-1}).
\end{equation}
 
\item \textbf{Computing the posterior.} Using the updated parameters $\beta_{B}$, we compute the posterior of the $B$ matrix and apply it from the next step onward.
\end{enumerate}

Through the above parameter updates, which can be regarded as Hebbian-like associative learning rules at the level of generative-model parameters\cite{da2020active}, the parent and infant agents learn sequentially from the outcomes of cooperative actions.
\section{Experiments}
 
To evaluate the proposed model, we conducted the following simulation experiments.
 
\subsection{Visceral State}
 
The infant's visceral state is represented as a $6 \times 6$ discrete grid world over a two-dimensional space of energy and body temperature~\cite{keramati2014homeostatic}. The coordinates $(x, y)$ of each cell are defined by
\begin{equation}
[a, b] := \{ x \in \mathbb{Z} \mid a \le x \le b\}, \qquad x, y \in [0, 5].
\end{equation}
For energy, $x = 0$ represents deficiency and $x = 5$ represents excess; for body temperature, $y = 0$ represents low body temperature and $y = 5$ represents high body temperature. The prior preference $C$ over visceral states is shown in Figure~\ref{fig:Cmatrix}. The agents maintain the state in the central region where $C$ is maximized through cooperative actions.
 
\subsection{Action Definition}\label{subsec:action_def}

The agents can choose one of five actions in the grid world. Each action and its corresponding state transition are described below.
 
\begin{enumerate}
\item \textbf{Cool}: A drop in body temperature due to energy consumption. Transition: $(x, y) \to (x - 1, y - 1)$.
 
\item \textbf{Warm}: A rise in body temperature due to energy consumption. Transition: $(x, y) \to (x - 1, y + 1)$.
 
\item \textbf{Eat}: An energy intake action. Transition: $x \to x + 2$, with body temperature $y$ changing with probability $20\%$. The direction of change is a further decrease when the body temperature is low ($y \le 2$) and a further increase when the body temperature is high ($y \ge 3$).
 
\item \textbf{Play}: A play action with energy consumption. Transition: $x \to x - 1$, with body temperature $y$ changing in the same manner as for Eat with probability $20\%$.
 
\item \textbf{Sleep}: A rest action. The energy state is unchanged ($x \to x$). With probability $20\%$, body temperature $y$ changes in the same manner as for Eat.
\end{enumerate}
 
We quantitatively evaluated the model's control performance under this environment and these actions.
 
\subsection{Experimental Setup}
 
In this experiment, to evaluate how the parent--infant model interaction affects the regulation of the infant's visceral state, we compared the following three conditions.
 
\begin{enumerate}
\item \textbf{A-led}: The infant always accepts the parent's proposal; conversely, when the infant's proposal differs from the parent's, the parent always rejects it.
 
\item \textbf{B-led}: The parent always accepts the infant's proposal; conversely, when the parent's proposal differs from the infant's, the infant always rejects it.
 
\item \textbf{MHNG}: parent--infant symbol mediated interaction using MHNG.
\end{enumerate}
Two cooperative actions (one trial each in the speaker and listener roles) constitute one iteration. We ran $10$ trials per condition with $1{,}000$ iterations.

\begin{figure}[t]
  \centering
  \begin{subfigure}[b]{0.5\linewidth}
    \centering
    \includegraphics[width=\linewidth]{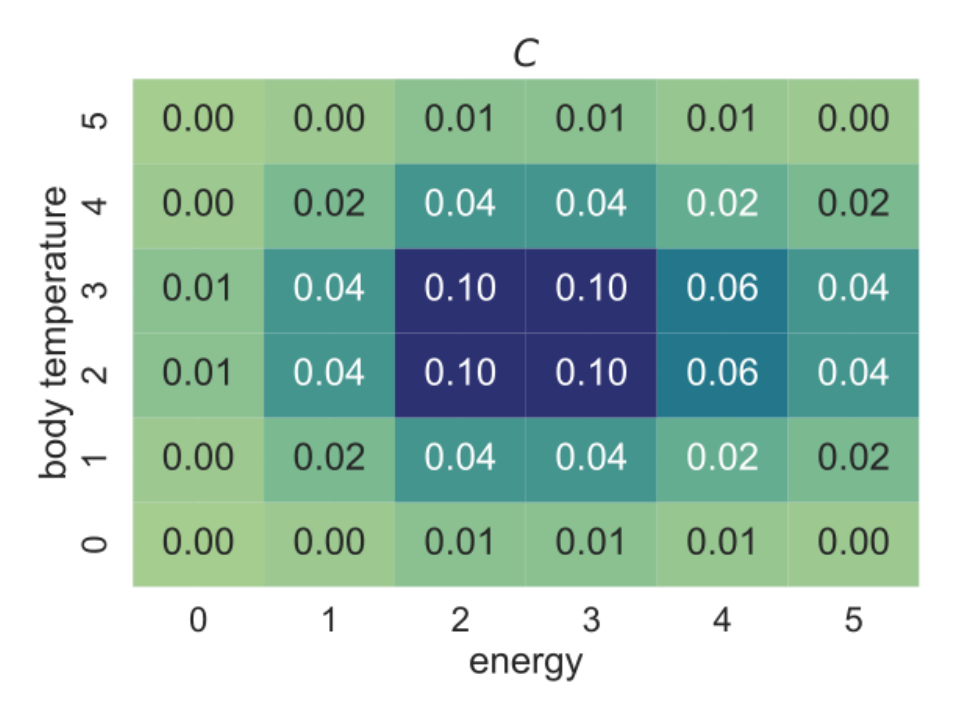}
    \caption{Setting of the prior preference $C$.}
    \label{fig:Cmatrix}
  \end{subfigure}
  \hfill
  \begin{subfigure}[b]{0.35\linewidth}
    \centering
    \includegraphics[width=\linewidth]{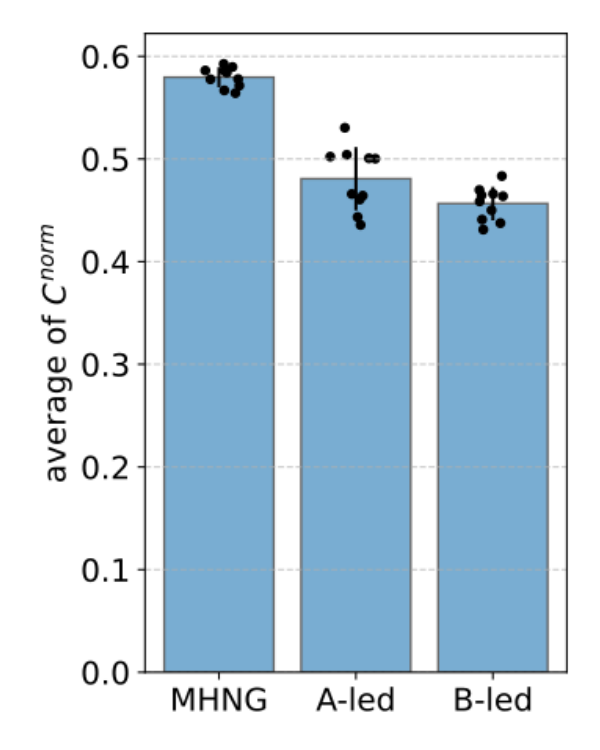}
    \caption{Comparison of the per-trial mean of $C^{\text{norm}}$ across conditions.}
    \label{fig:Cmean}
  \end{subfigure}
  \caption{The setting of $C$ and the result of $C^{norm}$ under each condition.}
\end{figure}

\subsubsection{Learning Metrics}
 
\paragraph{Adaptiveness of the visceral state $C^{\text{norm}}$.}
To quantify the extent to which the infant's visceral state is regulated toward regions of high prior preference $C$, we computed the adaptiveness at time $t$:
\begin{equation}
C^{\text{norm}}_{t} = \frac{C_{t}}{\max(C)},
\end{equation}
where $C_{t}$ is the value of the $C$ matrix at the cell corresponding to the visceral state at time $t$, and $\max(C)$ is the maximum value of the $C$ matrix. A value closer to $1$ indicates a closer match to the prior preference.

\paragraph{Generative-model learning error.}
To evaluate the learning progress of the generative matrices, we computed the Kullback--Leibler divergence between the true generative matrix and the matrix learned by each agent.
A decrease in KL divergence indicates that the learned matrix approaches the true generative matrix.
For the parent agent, we evaluated the sensory-generation matrix $A^{A}$ by comparing it with the true sensory-generation matrix $A^{true}$.
For each latent state $n_z$, the learning error of $A^{A}$ was defined as
\begin{equation}
D^{A}_{\mathrm{KL}}(t, n_z)=D_{\mathrm{KL}}\left[A^{\mathrm{true}}(\cdot \mid n_z)\,\middle\|\,A^{A}_{t}(\cdot \mid n_z)\right],
\end{equation}
and the average learning error was computed as
\begin{equation}
KLD[A^{true}||A^{A}](t)=\frac{1}{N_z}\sum_{n_z}D^{A}_{\mathrm{KL}}(t, n_z).
\end{equation}
For the infant agent, we evaluated the state-transition matrix $B^{B}$ for a given action $a$ by comparing it with the true transition matrix $B^{true}$.
For each previous latent state $n_z$, the learning error of $B^{B}$ was defined as
\begin{equation}
D^{B}_{\mathrm{KL}}(t, n_z; a)=D_{\mathrm{KL}}\left[B^{\mathrm{true}}(\cdot \mid n_z, a)\,\middle\|\,B^{B}_{t}(\cdot \mid n_z, a)\right],
\end{equation}
and the average learning error was computed as
\begin{equation}
KLD[B^{true}||B^{B}](t;a)=\frac{1}{N_z}\sum_{n_z}D^{B}_{\mathrm{KL}}(t, n_z; a).
\end{equation}
In the experiments, we evaluated the infant model using $B^{B}[\mathrm{Sleep}]$, corresponding to the sleep action.
 
\paragraph{latent representation similarity $\mathrm{JSD}^{z}$.}
We measure the similarity between the parent's and infant's latent representations $P(z^{A}_{t})$ and $P(z^{B}_{t})$ using the Jensen--Shannon divergence:
\begin{align}
\mathrm{JSD}^{z}_{t} &= \tfrac{1}{2} D_{\text{KL}}\!\left(P(z^{A}_{t}) \,\|\, M_{t}\right) + \tfrac{1}{2} D_{\text{KL}}\!\left(P(z^{B}_{t}) \,\|\, M_{t}\right), \\
M_{t} &= \tfrac{1}{2}\!\left(P(z^{A}_{t}) + P(z^{B}_{t})\right).
\end{align}
A value closer to $0$ indicates that the two latent representations are more similar.
We use latent-representation alignment to refer to a reduction of $\mathrm{JSD}^{z}_{t}$. 
In contrast, dynamic latent-representation synchrony is defined as sustained coupling of the agents’ latent-representation trajectories. 
This coupling is characterized by low time-integrated divergence and rapid re-alignment following stochastic state transitions.

\begin{figure}[!t]
  \centering
  \begin{subfigure}[b]{0.9\linewidth}
    \centering
    \includegraphics[width=\textwidth]{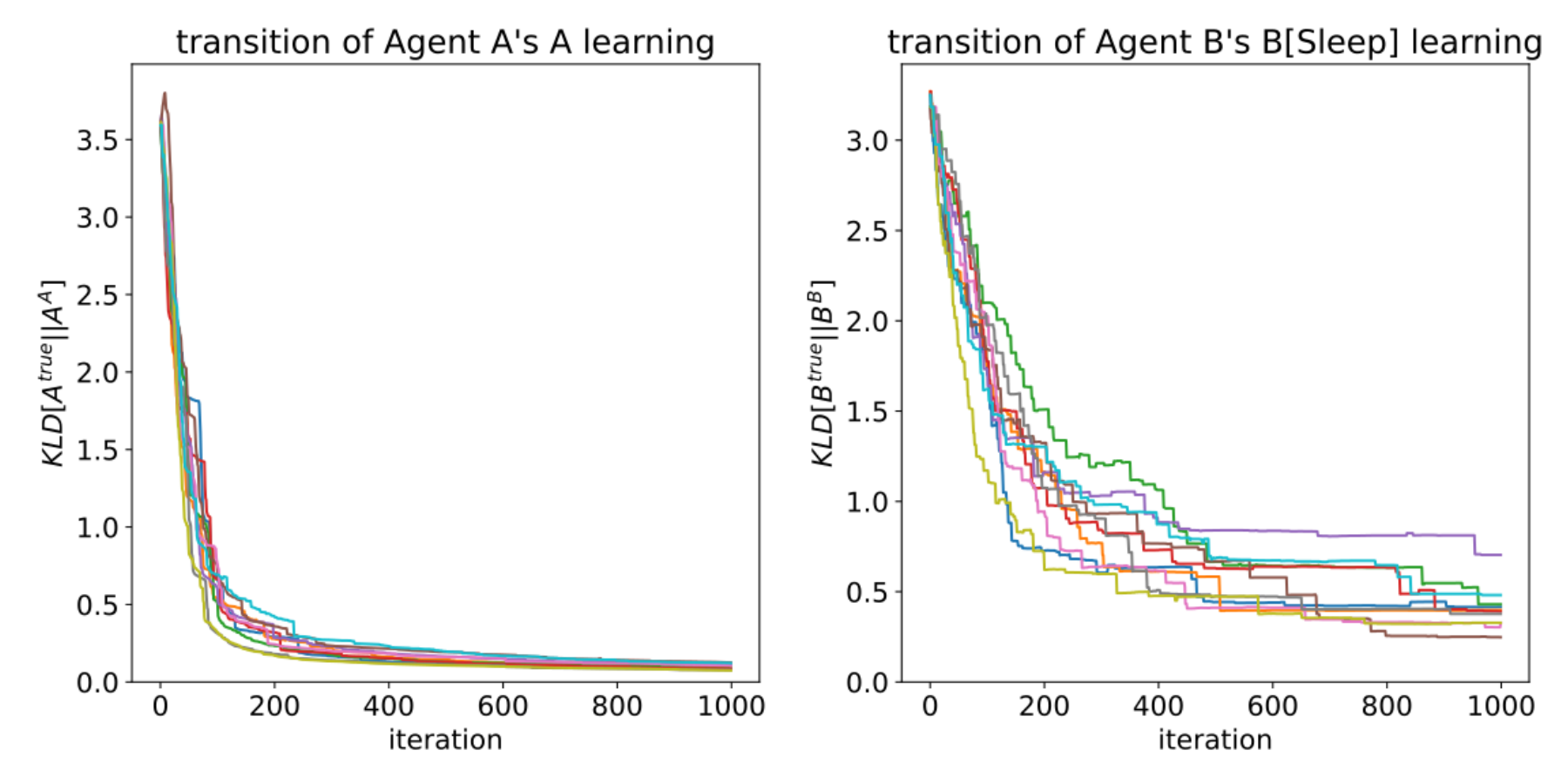}
    \caption{Evolution of KL divergence between true parameters and the parameters learned by each agent.}
    \label{fig:ABlearning}
  \end{subfigure}
  \hfill
  \begin{subfigure}[b]{0.6\linewidth}
    \centering
    \includegraphics[width=0.9\linewidth]{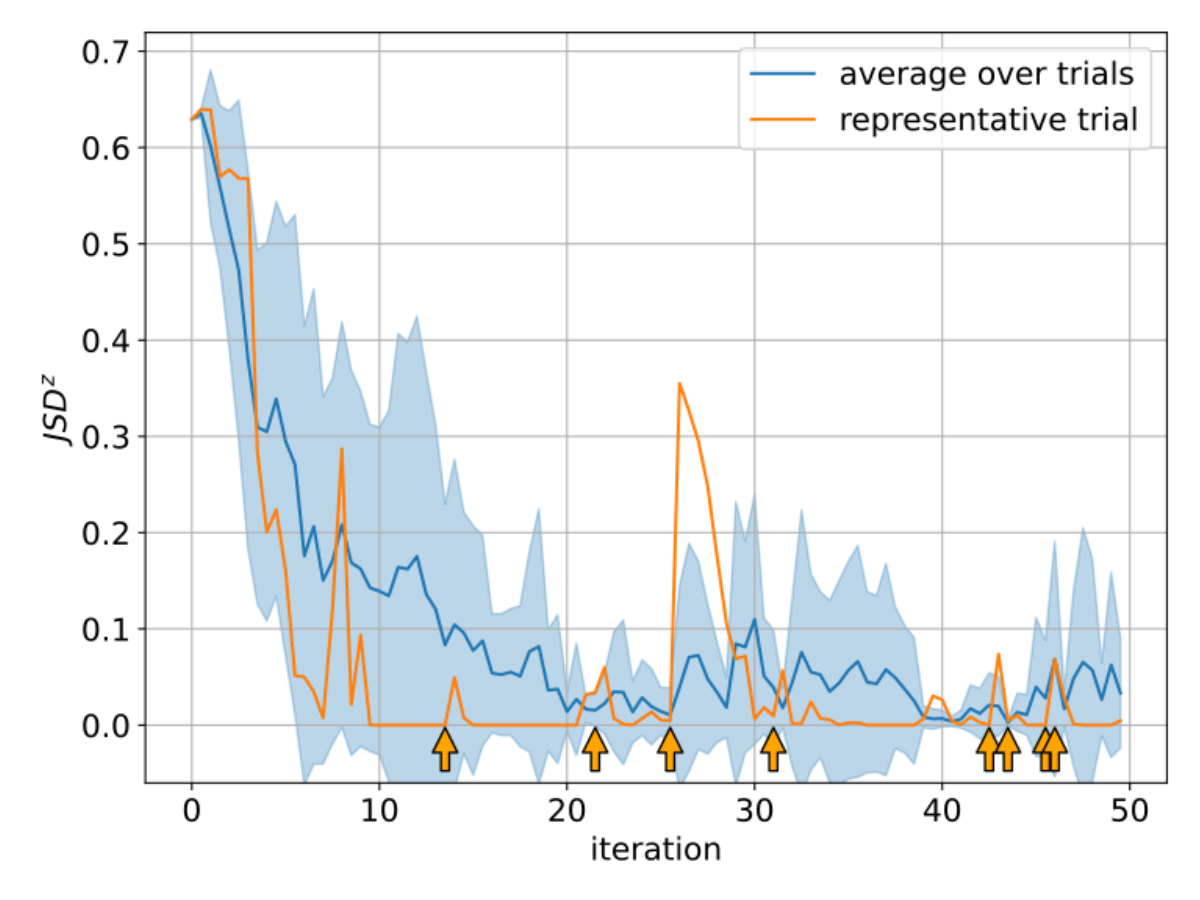}
    \caption{Onset and maintenance of latent-representation alignment measured by $\mathrm{JSD}^{z}$. Orange arrows indicate rare visceral transitions in the representative trial.}
    \label{fig:jsd_z}
  \end{subfigure}
    \hfill
  \begin{subfigure}[b]{0.3\linewidth}
    \centering
    \includegraphics[width=0.9\linewidth]{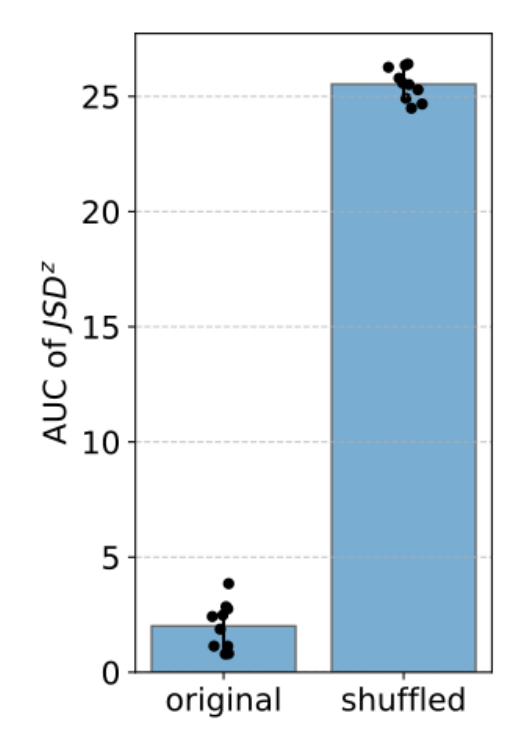}
    \caption{Area Under Curve (AUC) of $\mathrm{JSD}^{z}$ between the original interaction and a shuffled control in $20-50$ iterations}
    \label{fig:jsd_z_compare}
  \end{subfigure}
  \begin{subfigure}[b]{\linewidth}
    \centering
    \includegraphics[width=\linewidth]{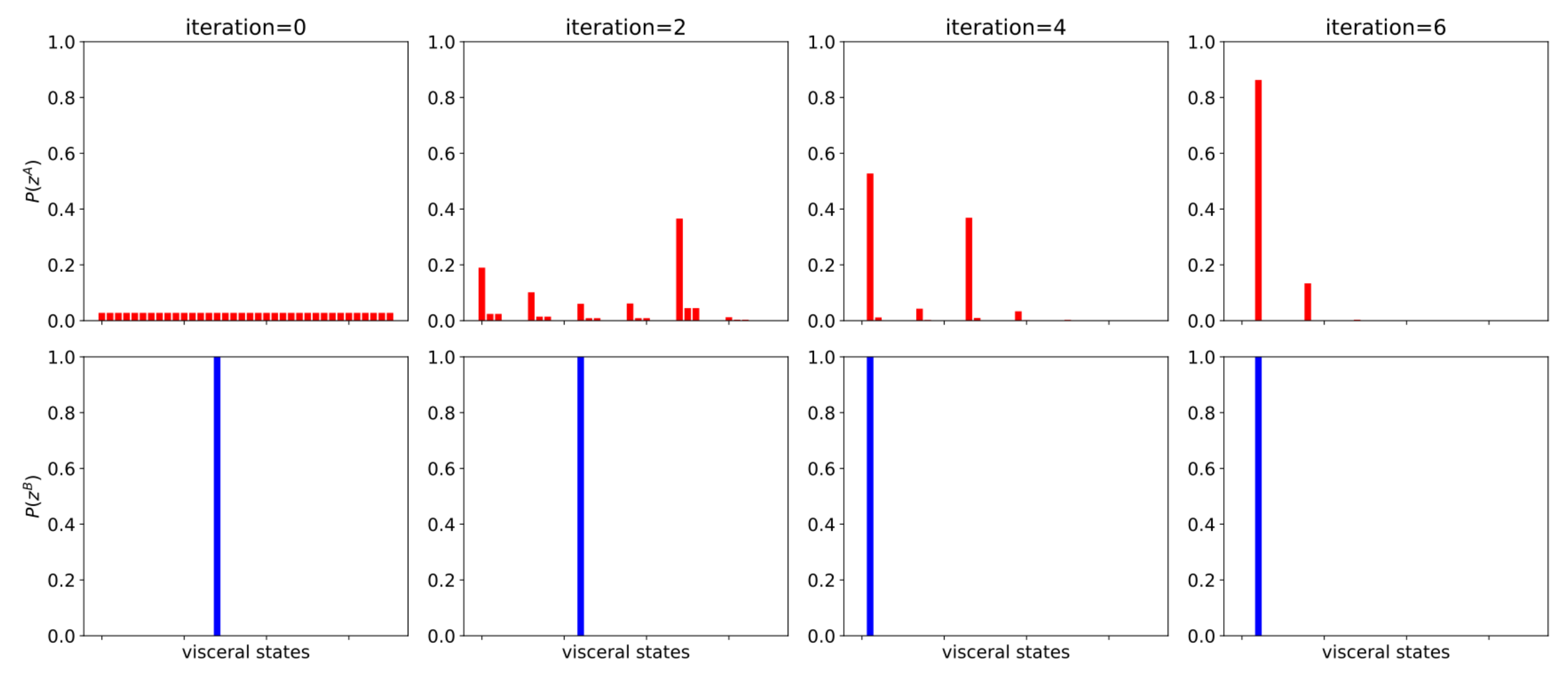}
    \caption{Representative onset of latent-representation alignment.}
    \label{fig:qs_transition}
  \end{subfigure}
  \caption{Generative-model learning and transition of latent representations under the MHNG condition.}
\end{figure}

\section{Results}
 
Under the MHNG condition, the infant's visceral state was maintained in the high prior preference region from the early phase of learning.
Figure~\ref{fig:Cmean} shows $C^{\text{norm}}$ averaged within each trial and aggregated by condition. The visceral state was kept at higher prior preference in the order MHNG, A-led, and B-led.

Figure~\ref{fig:ABlearning} shows trajectories of the generative-model learning errors for the two agents under MHNG condition: $\mathrm{KLD}[A^{\text{true}} \,\|\, A^A](t)$ for the parent's sensory-generation matrix and $\mathrm{KLD}[B^{\text{true}} \,\|\, B^B](t; a=\text{Sleep})$ for the infant's state-transition matrix corresponding to the sleep action.
Both quantities decreased monotonically and approached small values within approximately $500$ iterations, indicating that each agent's learnable matrix gradually converged toward the corresponding true generative matrix.

The initial evolution of the latent representation similarity $\mathrm{JSD}^{z}_{t}$ is shown in Figure~\ref{fig:jsd_z}. 
Note that the scale of the $x$-axis differs from that of Figure~\ref{fig:ABlearning}. 
After approximately $20$ iterations, $\mathrm{JSD}^{z}_{t}$ approached $0$, suggesting that the parent's and infant's latent representations had become aligned. 
It occurred far more rapidly than the decrease of KL divergence between parameters, meaning that the onset of latent-representation alignment preceded the learning of the generative model.
Figure~\ref{fig:qs_transition} suggests that this rapid alignment can be interpreted in terms of the asymmetric information processing implemented in the two agents' generative models.
The infant agent can infer its visceral state in a bottom-up manner through its accurate sensory-generation matrix $A^{B}$, corresponding to direct interoceptive perception. 
In contrast, the parent agent cannot directly access the infant's visceral state, but can infer it in a more top-down manner through its accurate state-transition matrix $B^{A}$, using the history of regulatory actions and their expected consequences. 
For example, if the infant has just eaten, the parent agent can predict that a low-energy or hunger-like state is unlikely even before the mapping from bodily cues to visceral states has been fully learned. 

After the initial alignment, $\mathrm{JSD}^{z}_{t}$ did not remain exactly zero. Instead, it stayed in a low-divergence regime with brief spike-like increases.
These transient increases were temporally associated with low-probability stochastic transitions of the visceral state shown by the orange arrows in Figure~\ref{fig:jsd_z}. 
When the body-temperature transition followed the $20\%$ branch (see Section~\ref{subsec:action_def}), the infant’s posterior reflected the realized interoceptive state, whereas the parent’s posterior initially favored the more probable $80\%$ transition. 
Thus, the spikes reflected rational transient desynchronization under asymmetric available sensory access.
In addition, $\mathrm{JSD}^{z}_{t}$ increased transiently after rare stochastic transitions but rapidly returned toward the pre-event low-divergence regime. 
To test whether the low $\mathrm{JSD}^{z}$ reflected temporally coordinated coupling, we compared the original $\mathrm{JSD}^{z}$ trajectory with a temporal-shuffle control, in which Agent B's latent representation sequence $P(z^B_t)$ was randomly permuted in time before computing $\mathrm{JSD}^{z}$ with Agent A's sequence $P(z^A_t)$. 
Figure~\ref{fig:jsd_z_compare} shows that the AUC of $\mathrm{JSD}^{z}$ in iterations $20-50$ was substantially larger in the shuffled condition than in the original condition. This result suggests that the original interaction maintained the agents in a low-$\mathrm{JSD}^{z}$ regime across successive state transitions.
We therefore characterize the observed phenomenon as dynamic latent-representation synchrony: the agents became aligned, were transiently desynchronized by stochastic visceral transitions, and rapidly re-aligned through interaction.
\section{Conclusion}

In this paper, we proposed a POMDP model that performs MHNG based on parent--infant homeostatic co-regulation. Compared with one--sided control conditions, the MHNG condition more adaptively maintained the infant's visceral state. 
In addition, the agents’ latent representations rapidly became aligned and remained dynamically coupled across action-induced state transitions. 
We refer to this sustained alignment as dynamic latent-representation synchrony.
Importantly, the present model does not directly simulate neural activity or anatomical connectivity. 
Nevertheless, it provides a computational analogue relevant to IBS reported in hyperscanning studies.

The present model uses a simplified visceral state represented as a two-dimensional discrete grid world. 
In addition, the visceral state, the interoceptive sensory signals perceived by the infant agent, and the exteroceptive sensory signals perceived by the parent agent all carry the same amount of information. 
Future work calls for modeling that can be aligned with real organisms through continuous, multi-dimensional representations and the use of more expressive generative models. 
Moreover, in this paper, the prior preference $C$ and the interpretation matrix $E$ are shared identically in advance and remain static. 
As a result, this work does not address the dynamics of symbol emergence, which is a central focus of CPC~\cite{taniguchi2024collective}. 
In the future, by introducing learning rules for the $C$ and $E$ matrices, we aim to model the emergence of social meaning.

Despite these limitations, we proposed a constructive model of parent--infant homeostatic co-regulation and provide a minimal constructive account of how latent-representation synchrony can emerge through local interaction between agents with asymmetric knowledge and only locally available sensory access.
This account is compatible with IBS reported in hyperscanning studies and supports CPC as a candidate computational framework for understanding dynamic representational synchrony in dyadic interaction.

\begin{ack}
This work was supported by JSPS KAKENHI Grant Number JP23H04834.
\end{ack}

\bibliographystyle{plain}
\bibliography{reference}

\end{document}